\documentstyle[aps,psfig]{revtex}
\title{\bf Iron under Earth's core conditions: Liquid-state \\
thermodynamics and high-pressure melting curve \\
from {\em ab initio} calculations} 
\author{D. Alf\`{e}$^{1,2}$, G. D. Price$^1$ and M. J. Gillan$^2$
\smallskip \\
$^1$Research School of Geological and Geophysical Sciences \\
Birkbeck and University College London \\
Gower Street, London WC1E~6BT, UK 
\smallskip \\
$^2$Physics and Astronomy Department, University College London \\
Gower Street, London WC1E~6BT, UK}
\begin{document}
\maketitle
\begin{abstract}
{\em Ab initio} techniques based on density functional
theory in the projector-augmented-wave implementation are
used to calculate the free energy and a range of other
thermodynamic properties of liquid iron at high pressures
and temperatures relevant to the Earth's core. The {\em ab initio}
free energy is obtained by using thermodynamic integration
to calculate the change of free energy on going from a
simple reference system to the {\em ab initio} system,
with thermal averages computed by {\em ab initio}
molecular dynamics simulation. The reference system
consists of the inverse-power pair-potential model used in
previous work. The liquid-state free energy is combined with
the free energy of hexagonal close packed Fe calculated earlier
using identical {\em ab initio} techniques to  obtain the
melting curve and volume and entropy of melting. Comparisons
of the calculated melting properties with experimental measurement
and with other recent {\em ab initio} predictions are presented.
Experiment-theory comparisons are also presented for the pressures
at which the solid and liquid Hugoniot curves cross the
melting line, and the sound speed and Gr\"{u}neisen parameter along
the Hugoniot. Additional comparisons are made with a commonly used equation
of state for high-pressure/high-temperature Fe based on experimental
data.
\end{abstract}

\section{Introduction}
\label{sec:intro}

The last few years have seen important progress in calculating the
thermodynamic properties of condensed matter using {\em ab initio}
techniques based on density-functional theory
(DFT)~\cite{sugino95,smargiassi95a,dewijs98,alfe99a}. There has been particular
attention to the thermodynamics of crystals, whose harmonic free
energy can be obtained from phonon frequencies computed by standard
DFT
methods~\cite{karki00,Lichtenstein00,xie99a,xie99b,lazzeri98,pavone98,alfe01}. The
{\em ab initio} treatment of liquid-state thermodynamics is also
important, and thermodynamic integration has been shown to be an
effective way of calculating the DFT free energy of
liquids~\cite{sugino95,dewijs98,alfe99a}. These developments have made
it possible to treat phase equilibria, including melting properties,
by completely {\em ab initio} methods. We report here DFT free-energy
calculations on high-pressure/high-temperature liquid iron, which we
combine with earlier results on the solid~\cite{alfe01} 
to obtain the complete
melting curve and the variation of the volume and entropy of melting
along the curve. We also present results for some key thermodynamic
properties of the liquid, which we compare with data from shock
experiments and other sources. The general methods developed here may
be useful for other problems involving phase equilibria under extreme
conditions. A brief report of this work was presented earlier~\cite{alfe99d}.

The properties of high-pressure/high-temperature Fe are of great
scientific importance because the Earth's core consists mainly of Fe,
with a minor fraction of light
impurities~\cite{birch64,ringwood77,poirier94}. The melting curve is
particularly important, since it provides one of the very few ways of
estimating the temperature at the boundary between the liquid outer
core and the solid inner 
core~\cite{anderson97}. 
Because of this, strenuous
efforts have been made to measure the melting
curve~\cite{boehler93,saxena94,shen98,errandonea01,williams87,yoo93,brown86},
but the extreme pressures and temperatures required ($p \sim 330$~GPa,
$T \sim 6000$~K) make the experiments very demanding. {\em Ab initio}
calculations therefore have a major role to play, and several
independent attempts to obtain the melting curve using different {\em
ab initio} strategies have been reported
recently~\cite{alfe99d,laio00,belonoshko00}. The rather unsatisfactory
agreement between the predictions makes a full presentation of the
technical methods all the more important.

The calculation of melting properties using {\em ab initio} free
energies was pioneered by Sugino and Car~\cite{sugino95} in their work
on the melting of Si at ambient pressure.  Related methods were
subsequently used by de~Wijs {\em et al.}~\cite{dewijs98} to study the
melting of Al. In both cases, thermodynamic integration (see
e.g. Ref.~\cite{frenkel96}) was used to obtain the {\em ab initio}
free energy from the free energy of a simple reference system, and we
follow the same strategy here. The other recent
calculations~\cite{laio00,belonoshko00} on the high-pressure melting
of Fe employed {\em ab initio} methods in a different way. Free
energies were not calculated, but instead an empirical parameterised
form of the total-energy function was fitted to DFT total energies
calculated for representative configurations of the solid and liquid.
The empirical energy function was then used in molecular dynamics
simulations of very large systems containing coexisting solid and
liquid.

The detailed DFT techniques used in this work are identical to those
used in our work on h.c.p. Fe~\cite{alfe01}. In particular, we use the
generalised gradient approximation (GGA) for exchange-correlation
energy, in the form known as Perdew-Wang 1991~\cite{wang91,perdew92},
which reproduces very accurately a wide range of experimental
properties of solid iron, as noted in more detail
elsewhere~\cite{stixrude94,soderlind96,vocadlo97,alfe99b}. We also use the
projector-augmented-wave (PAW) implementation of
DFT~\cite{alfe99b,blochl94,kresse99}, which is an all-electron
technique similar to other standard implementations such as
full-potential augmented plane waves (FLAPW)~\cite{wei85}, 
as well as being closely
related to the ultrasoft pseudopotential method~\cite{vanderbilt90}.
We have used the VASP code~\cite{kresse96a,kresse96b}, which is
exceptionally stable an efficient for metals, with the implementation
of an extrapolation of the charge density which increases the
efficiency of molecular dynamics simulations by almost a factor of
two~\cite{alfe99e}.

The calculation of melting properties demands very high precision for
the free energies of the two phases, as emphasised 
elsewhere~\cite{dewijs98,alfe01}. The
required precision is set by the value of the entropy of melting, and
one finds that in order to calculate the melting temperature to within
100~K the non-cancelling error in the free energies must be reduced to
$\sim 10$~meV/atom.  The use of identical electronic-structure methods
in the two phases is clearly necessary; but it is certainly not
sufficient, since the detailed free-energy techniques differ in the
two phases.  In the solid, we relied heavily on harmonic calculations,
whereas the liquid-state calculations rely on relating the free energy
to that of a reference liquid. It is therefore essential to reduce the
statistical-mechanical errors below the tolerance, and we aim to
demonstrate that this has been achieved.

In the next Section, we summarise the technical methods, and Sec.~3
then reports our results for the DFT free energy of liquid Fe over a
wide range of thermodynamic states. Sec.~4 presents our calculated
melting properties, which we compare with experimental results and the
predictions of other {\em ab initio} calculations. Our free-energy
results have been used to compute a variety of other thermodynamic
quantities for the liquid, and we compare these in
Secs.~\ref{sec:hugoniot} and \ref{sec:liq_therm} with direct shock
measurements as well as published extrapolations of other experimental
data. In the final Section, we give further discussion and a summary
of our conclusions.

\section{Techniques}
\label{sec:techniques}

The key thermodynamic quantity calculated in this work is the {\em ab
initio} Helmholtz free energy $F$, which, with the statistical
mechanics of the nuclei treated in the classical limit, is:
\begin{equation}
F = - k_{\rm B} T \ln \left\{
\frac{1}{N ! \Lambda^{3 N}} \int d {\bf R}_1 \ldots d {\bf R}_N \,
\exp \left[ - \beta U_{\rm AI} ( {\bf R}_1 , \ldots {\bf R}_N ;
T_{\rm el} ) \right] \right\} \; ,
\end{equation}
where ${\bf R}_i$ ($i = 1 , \ldots N$) are the positions of the $N$
nuclei, $\Lambda = h / ( 2 \pi M k_{\rm B} T )^{1/2}$ is the thermal
wavelength, with $M$ the nuclear mass and $\beta = 1 / k_{\rm B}
T$. The quantity $U_{\rm AI} ( {\bf R}_1 , \ldots {\bf R}_N ; T_{\rm
el} )$ is the DFT electronic free energy calculated with the $N$
nuclei fixed at positions ${\bf R}_1 , \ldots {\bf R}_N$. This is
given, following the Mermin formulation of finite-temperature 
DFT~\cite{mermin65}, by
$U_{\rm AI} = E - T S$, where the DFT energy $E$ is the usual sum of
kinetic, electron-nucleus, Hartree and exchange-correlation terms, and
$S$ is the electronic entropy, given by the independent-electron
formula: $S = - k_{\rm B} T_{\rm el} \sum_i [ f_i \ln f_i + ( 1 - f_i
) \ln ( 1 - f_i ) ]$, with $f_i$ the thermal (Fermi-Dirac) occupation
number of orbital $i$. In exact DFT, the exchange correlation (free)
energy $E_{\rm xc}$ has an explicit dependence on $T_{\rm el}$ but we
assume here that $E_{\rm exc}$ has its zero-temperature form. It was
shown in Ref.~\cite{alfe01} that quantum corrections to the
classical approximation to the free energy are negligible
in the high-temperature solid, and the same will be true
{\em a fortiori} in the liquid.

Our earlier work~\cite{alfe99b} should be consulted for technical
details of the PAW implementation of DFT that we use. We note here
that under Earth's core conditions it is not accurate enough to
neglect the response of the $3p$ electrons, and in principle these
should be explicitly included in the valence set along with the $3d$
and the $4s$ electrons. However, as shown
earlier~\cite{vocadlo97,alfe99b}, the computational effort of
including 3$p$ electrons explicitly can be avoided with almost no loss of
precision by mimicking the effect of the 3$p$ electrons by an
effective pair potential.  The procedure for constructing this pair
potential was described in Ref.~\cite{alfe99b}. The pair potential
used here is exactly the same as we used in our thermodynamic
calculations on the h.c.p. solid, so that a good cancellation of any
residual errors is expected. To avoid any possible doubt on this
score, we have also done spot checks on the effect of including both
3$s$ and 3$p$ electrons explicitly in the valence set, as reported in
Sec.~\ref{sec:melting}.  The outermost core radius in our PAW
calculations is 1.16~\AA. At Earth's core pressures, the atoms in both
liquid and solid come closer than the diameter of the ionic cores,
which therefore overlap. We will show in Sec.~\ref{sec:melting} that
this too has only a very small effect on the free energies.

The present calculations, like those reported earlier on the h.c.p.
solid, make use of `thermodynamic integration'~\cite{frenkel96}, which
is a completely general technique for determining the difference of
free energies $F_1 - F_0$ of two systems whose total-energy functions
are $U_1$ and $U_0$. The basic idea is that $F_1 - F_0$ represents the
reversible work done on continuously and isothermally switching the
energy function from $U_0$ to $U_1$. To do this switching, a
continuously variable energy function $U_\lambda$ is defined as:
\begin{equation}
U_\lambda = ( 1 - \lambda ) U_0 + \lambda U_1 \; ,
\label{eqn:thermint_1}
\end{equation}
so that the energy goes from $U_0$ to $U_1$ as $\lambda$
goes from 0 to 1. In classical statistical mechanics, the work
done in an infinitesimal change $d \lambda$ is:
\begin{equation}
d F = \langle d U_\lambda / d \lambda \rangle_\lambda d \lambda =
\langle U_1 - U_0 \rangle_\lambda d \lambda \; ,
\label{eqn:thermint_2}
\end{equation}
where $\langle \, \cdot \, \rangle_\lambda$ represents the
thermal average evaluated for the system governed by $U_\lambda$.
It follows that:
\begin{equation}
F_1 - F_0 = \int_0^1 d \lambda \, \langle U_1 - U_0 \rangle_\lambda \, .
\label{eqn:thermint_3}
\end{equation}
We use this technique to calculate the {\em ab initio} free energy
$F_{\rm AI}$ of liquid Fe by identifying $U_1$ as the {\em ab initio}
total energy function $U_{\rm AI} ( {\bf R}_1 , \ldots {\bf R}_N )$
and $U_0$ as the total energy $U_{\rm ref} ( {\bf R}_1 , \ldots {\bf
R}_N )$ of a simple model reference system, whose free energy $F_{\rm
ref}$ can be calculated. Then $F_{\rm AI}$ is given by:
\begin{equation}
F_{\rm AI} = F_{\rm ref} + \int_0^1 d \lambda \,
\langle U_{\rm AI} - U_{\rm ref} \rangle_\lambda \; .
\end{equation}
In practice, we calculate $\langle U_{\rm AI} - U_{\rm ref}
\rangle_\lambda$ for a suitable set of $\lambda$ values, and perform
the integration numerically.  The average $\langle U_{\rm AI} - U_{\rm
ref} \rangle_\lambda$ is evaluated at each $\lambda$ using
constant-temperature {\em ab initio} molecular dynamics, with the time
evolution generated by the total-energy function $U_\lambda = ( 1 -
\lambda ) U_{\rm ref} + \lambda U_{\rm AI}$.

As explained in more detail elsewhere~\cite{alfe01}, the computational 
effort needed to perform the thermodynamic
integration is greatly reduced if the fluctuations of the energy
difference $\Delta U \equiv U_{\rm AI} - U_{\rm ref}$ are small, for
two reasons. First, the amount of sampling needed to calculated
$\langle \Delta U \rangle_\lambda$ to a given precision is reduced;
second, the variation of $\langle \Delta U \rangle_\lambda$ as
$\lambda$ goes from 0 to 1 is reduced. In fact, if the fluctuations
$\delta \Delta U \equiv \Delta U - \langle \Delta U \rangle_{\rm AI}$
are small enough, one can neglect this variation and approximate
$F_{\rm AI} \simeq F_{\rm ref} + \langle \Delta U \rangle_{\rm AI}$,
with the average taken in the {\em ab-initio} ensemble. If this is not
good enough, the next approximation is readily shown to be:
\begin{equation}\label{eqn:second_order}
F_{\rm AI} \simeq F_{\rm ref} + \langle \Delta U \rangle_{\rm AI} +
\frac{1}{2 k_{\rm B} T} \langle ( \delta \Delta U )^2 \rangle_{\rm AI}
\; .
\label{eqn:2nd_order}
\end{equation}
Our task is therefore to search for a model $U_{\rm ref}$ for which
the fluctuations $U_{\rm AI} - U_{\rm ref}$ are as small as possible.

The problem of mimicking the fluctuations of {\em ab initio} energy
$U_{\rm AI}$ in high-$p$/high-$T$ liquid Fe using a reference system
was studied in detail in a recent paper~\cite{alfe99b}. We showed there that a
$U_{\rm ref}$ consisting of a sum of pair potentials:
\begin{equation}
U_{\rm ref}  =  U_{\rm th} + U_{\rm pair} \; ,
\label{eqn:pair}
\end{equation}
in which
\begin{equation}
U_{\rm pair}  =  \frac{1}{2} \sum_{i \ne j}
\phi ( \mid {\bf R}_i - {\bf R}_j \mid ) \; ,
\end{equation}
can be arranged to mimic the fluctuations of $U_{\rm AI}$
very precisely, if we choose $\phi ( r )$ to be a repulsive
inverse-power potential $\phi ( r ) = B / r^\alpha$, with suitable
values of $B$ and $\alpha$.
In this expression for $U_{\rm ref}$, we have
included a term $U_{\rm th}$ which depends on thermodynamic state, but
does not depend on the positions ${\bf R}_i$. We define this $U_{\rm
th}$ so as to minimize the mean square value of $\Delta U$, which
requires the following condition:
\begin{equation}
\langle \Delta U \rangle_{\rm AI} \equiv
\langle U_{\rm AI} - U_{\rm pair} - U_{\rm th} \rangle_{\rm AI} = 0 \; .
\label{eqn:Uth_def}
\end{equation}
It might at
first seem surprising that such a simple $U_{\rm ref}$ can reproduce
$U_{\rm AI}$ accurately, since it does not explicitly represent the
metallic bonding due to partial filling of the $3 d$-band, which
is absorbed into the term $U_{\rm th}$. However,
the only requirement on $U_{\rm ref}$ is that the fluctuations $\delta
\Delta U$ should be small for the given liquid state, and we shall
demonstrate below that this is indeed achieved. We comment further on this
question in Sec.~\ref{sec:discussion}.

The free energy of the reference system can be expressed as:
\begin{equation}
F_{\rm ref} = U_{\rm th} + F_{\rm pair} \; ,
\end{equation}
where $F_{\rm pair}$ is the free energy associated with $U_{\rm
pair}$.  An important advantage of our chosen form of $U_{\rm ref}$ is
that $F_{\rm pair}$ depends non-trivially only on a single
thermodynamic variable, rather than depending separately on
temperature $T$ and atomic number density $n$. Let $F_{\rm pair}^x
\equiv F_{\rm pair} - F_{\rm pg}$ be the excess free energy of the
reference system, {\em i.e.} the difference between $F_{\rm pair}$ and
the free energy $F_{\rm pg}$
of the perfect gas at the given $T$ and $n$. Then the
quantity $f_{\rm pair}^x \equiv F_{\rm pair}^x / N k_{\rm B} T$
depends only on the dimensionless thermodynamic parameter $\zeta$
defined as:
\begin{equation}
\zeta = B n^{\alpha / 3} / k_{\rm B} T \; .
\end{equation}
This simplifies the representation of $F_{\rm ref}$,
since we can write:
\begin{equation}
F_{\rm ref} = U_{\rm th} + F_{\rm pg} +
N k_{\rm B} T f_{\rm pair}^x ( \zeta ) \; .
\label{eqn:F_ref}
\end{equation}

We also expect the representation of $U_{\rm th}$ to be simple.
Note first that since $U_{\rm th}$ is defined so that
$\langle \Delta U \rangle_{\rm AI} =0$,
then from Eqn~(\ref{eqn:Uth_def})
we must have:
\begin{equation}
U_{\rm th} = \langle U_{\rm AI} - U_{\rm pair} \rangle_{\rm AI} \; .
\end{equation}
Now if the fluctuations of $U_{\rm AI} - U_{\rm pair}$ are indeed
small, then the value of $\langle U_{\rm AI} - 
U_{\rm pair} \rangle_{\rm AI}$ should be very close to the value
of $U_{\rm AI} - U_{\rm pair}$ evaluated at zero temperature
with all atoms on the sites of the perfect h.c.p. lattice having
the same density as the liquid. Denoting by $U_{\rm AI}^0$
and $U_{\rm pair}^0$ the values of the zero-temperature
{\em ab initio} and pair-potential energies for the perfect
lattice, and defining $U_{\rm th}^0 \equiv U_{\rm AI}^0 -
U_{\rm pair}^0$, we can then write:
\begin{equation}
U_{\rm th} = U_{\rm th}^0 + \delta U_{\rm th} \; ,
\label{eqn:delta_Uth}
\end{equation}
where $\delta U_{\rm th}$ will be a small quantity depending
weakly on volume and temperature. The accurate computation
and representation of $U_{\rm AI}^0$ were discussed in 
Ref.~\cite{alfe01},
and the accurate computation of $U_{\rm pair}^0$ is clearly
trivial, so that the treatment of $U_{\rm th}^0$ is straightforward.
The small difference $\delta U_{\rm th} \equiv \langle U_{\rm AI} -
U_{\rm pair} \rangle_{\rm AI} - U_{\rm th}^0$
is evaluated from the AIMD simulations described in 
Sec.~\ref{sec:ref2full}.

We conclude this Section by summarising our route to the
calculation of the {\em ab initio} free energy $F_{\rm AI}$
of the liquid. We employ Eqn~(\ref{eqn:2nd_order}), 
ignoring the higher-order
fluctuation terms. Recalling that $\langle \Delta U \rangle_{\rm AI} = 0$,
and combining Eqns~(\ref{eqn:F_ref}) 
and (\ref{eqn:delta_Uth}), we have:
\begin{eqnarray}
F_{\rm AI} & \simeq & F_{\rm ref} + 
\langle ( \delta \Delta U )^2 \rangle_{\rm AI} / 2 k_{\rm B} T
\nonumber \\
           & = & U_{\rm th}^0 + \delta U_{\rm th} + F_{\rm pg} +
N k_{\rm B} T f_{\rm pair}^x ( \zeta ) +
\langle ( \delta \Delta U )^2 \rangle_{\rm AI} / 2 k_{\rm B} T \; .
\label{eqn:F_AI}
\end{eqnarray}
We now turn to the calculation of the reduced free energy
$f_{\rm pair}^x ( \zeta )$ of the reference system and the small
quantities $\delta U_{\rm th}$ and $\langle ( \delta \Delta U )^2
\rangle_{\rm AI}$. We shall also give evidence that with our
chosen reference model the higher-order fluctuation terms
omitted from Eqn~(\ref{eqn:F_AI}) are indeed negligible.

\section{Free energy of the liquid}
\label{sec:free_energy}

\subsection{Inverse-power reference system}
\label{sec:ip_fit}
The PAW calculations used to validate the inverse-power
reference system are those reported in Ref.~\cite{alfe99b}. They
consist of a set of AIMD simulations performed at 16 thermodynamic
states covering the temperature range $3000 - 8000$~K and the
pressure range $60 - 390$~GPa. All the simulations were
performed on a 67-atom system using $\Gamma$-point sampling,
with a time step of 1~fs. We stress that such a small system
with such limited sampling cannot be expected to yield very precise
results for thermodynamic quantities, and our only purpose
here is to demonstrate the adequacy of the reference system.
At each thermodynamic state, the system was equilibrated using the
reference system itself, and AIMD data were then
accumulated for a time span of 5~ps.

We showed in Ref.~\cite{alfe99b} that the inverse-power model, with
parameters $\alpha = 5.86$ and $B$ chosen so that for $r = 2$~\AA\ the
potential $\phi (r)$ is 1.95~eV, reproduces very closely the {\em ab
initio} liquid for the state $T = 4300$~K, $\rho =
10700$~kg~m$^{-3}$. We have studied the strength of the $\delta \Delta
U$ fluctuations for all 16 thermodynamic states, using exactly the
same reference model for all states, and we report in
Table~\ref{tab:fluct} the normalized strength of these
fluctuations, which we characterize by the quantity $\sigma \equiv
\left[ \langle ( \delta \Delta U )^2 \rangle_{\rm AI} / N
\right]^{1/2}$. Two points should be noted: First, $\sigma$ is small,
since its typical value of 100~meV is markedly smaller than the
typical thermal energies $k_{\rm B} T$ (258~meV at the lowest
temperature of 3000~K). Once $\sigma$
is as small as this, little is gained by further improvement of the
reference system. Second, $\sigma$ does not vary strongly with
thermodynamic state, so that the reference system specified
by the values of $\alpha$ and $B$ given above can be used for
all the thermodynamic states of interest here.

\subsection{Free energy of reference system}
\label{sec:ip_free}
In order to cover the range of thermodynamic states of liquid Fe that
interests us, we need accurate values of the excess free energy of the
reference system for $\zeta$ values going from 2.5 to 5.0.  There have
been many studies of the thermodynamic properties of inverse-power
systems, including one on the free energy of the liquid $1 / r^6$
system~\cite{laird92}, but since these do not provide what we need we
have made our own calculations of $f_{\rm ref}^x ( \zeta )$ for the $1
/ r^{5.86}$ case. Our strategy is to start with standard literature
values for the excess free energy of the Lennard-Jones (LJ) liquid (we
use the results reported in Ref.~\cite{johnson93}), and to use
thermodynamic integration to go from the LJ system to the
inverse-power system, so that $U_0$ and $U_1$ in 
Eqns~(\ref{eqn:thermint_1}--\ref{eqn:thermint_2})
represent the LJ and inverse-power total energies respectively. In
doing this, our target was to keep technical errors small enough so
that the final free energy $F_{\rm ref}$ is correct to better than
5~meV/atom.

We note the following technical points. The calculations were done at a
standard volume per atom, usually taken to be 8.67~\AA$^3$, with the
temperature chosen to give the required value of $\zeta$. Ewald
techniques were used to avoid cutting off the inverse-power potential
at any distance -- we regarded this as essential, since a cut-off
would compromise the scaling properties of the reference system. The
classical molecular dynamics simulations used to compute $\langle U_1
- U_0 \rangle_\lambda$ were done using the constant-temperature
technique, with each atom taken to have a mass of 55.86 a.u., and the
time-step set equal to 1~fs. For each thermodynamic state, we are free
to choose any convenient values for the LJ parameters $\epsilon$ and
$\sigma$.  Our criterion for choosing these is that the fluctuations
of $U_1 - U_0$ should be kept reasonably small, but with the proviso
that the initial LJ system must be in the liquid state. In many cases,
we have checked for consistency by using different $\epsilon$ and
$\sigma$ values. Since we require $f_{\rm ref}^x ( \zeta )$ in the
thermodynamic limit of infinite system size, we have made careful
checks on size effects. Tests on systems containing up to 499 atoms
show that size errors in $f_{\rm ref}^x ( \zeta )$ are less than 1
meV/atom, and this is small enough to ensure that $F_{\rm ref}$ has a
precision of better than 5~meV.  Most of this error arises from the
error in the literature values of the LJ free energy. As a further
check on our techniques, we have done calculations on the $1 / r^6$
system at selected thermodynamic states, and compared with the free
energy results of Laird and Haymet~\cite{laird92}.

We have checked our procedures by repeating most of the
calculations using the perfect gas as reference system, so as to be
free from possible errors in the free energy of the LJ system. For
these calculations we used a different form for $U_\lambda$, namely
\begin{equation}
U_\lambda = ( 1 - \lambda^2 ) U_0 + \lambda^2 U_1 \; ,
\end{equation}
where $U_1$ is the total energy of the inverse power system and $U_0 =
0$ is that of the perfect gas. Using this functional form 
Eq.~(\ref{eqn:thermint_3}) becomes
\begin{equation}
F_1 - F_0 = \int_0^1 d \lambda \, 2\lambda \langle U_1 - U_0 \rangle_\lambda \, .
\end{equation}
The advantage of using this different functional form for $U_\lambda$
is that the value of the integrand does not need to be computed for
$\lambda = 0$, where the dynamics of the system is that of the
perfect gas. In this case, since there are no forces in the
system, there is nothing to prevent the atoms from overlapping, and the
potential energy $U_1$ diverges. Not computing the integrand at
$\lambda=0$ only partially solves this problem, since for small values
of $\lambda$ the forces on the atoms are small, the atoms can come
close together, and the potential energy $U_1$ fluctuates
violently. However, we found that by performing long enough
simulations, typically 1~ns, we could calculate the integral with an
accuracy of {\em ca.}~1~meV/atom. These calculations with the
perfect-gas reference system give excess free energies of the inverse
power system that are systematically 5~meV/atom lower than
those obtained using the LJ reference system. Our belief is
that the discrepancy arises from a small systematic 
error in the free energies given in Ref.~\cite{johnson93}.

After all these tests, calculations of $f_{\rm ref}^x ( \zeta )$ were
done at a regularly spaced set of $\zeta$ values at intervals of 0.25,
and we found that the results could be fitted to the required
precision by the following 3rd-degree polynomial:
\begin{equation}
f_{\rm ref}^x ( \zeta ) = \sum_{i=0}^3 c_i \zeta^i \; .
\end{equation}
The values of the coefficients are: $c_0=1.981, c_1=5.097,
c_2=0.1626, c_3 = 0.009733$.

\subsection{From reference to full {\em ab initio}}
\label{sec:ref2full}

To achieve our target precision of 10~meV/atom in the {\em ab initio}
free energy $F_{\rm AI}$ of the liquid,
two sources of error must
be studied: system size effects and electronic $k$-point sampling.
An important point to note is that these errors only affect the small
terms $\delta U_{\rm th}$ and $\langle ( \delta \Delta U )^2
\rangle_{\rm AI} / 2 k_{\rm B} T$ in Eqn~(\ref{eqn:F_AI}), since 
$f_{\rm ref}^x ( \zeta )$ refers already to the infinite
system, and $k$-point errors in $U_{\rm th}^0$ are negligible.
We also study the validity of neglecting the higher-order
fluctuation terms in Eqn~(\ref{eqn:F_AI}).

We focus first on the quantity $\delta U_{\rm th}$ in
Eqn~(\ref{eqn:F_AI}). To study size errors in this quantity, 
we calculated the thermal average $\langle
U_{\rm AI} - U_{\rm pair} \rangle_{\rm AI}$ for a range of system
sizes. These test calculations were done on systems
of up to 241 atoms at $V = 8.67$~\AA$^3$/atom and $T = 4300$~K using
$\Gamma$-point sampling. The preparation and equilibration of these
systems were done using the inverse-power reference system. Since the
latter so closely mimics the {\em ab initio} system for the 67-atom
cell, it should provide a well equilibrated starting point for {\em ab
initio} simulation of larger systems. The duration of all the {\em ab
initio} simulations after equilibration was 1~ps. The results of these
tests are summarised in Table~\ref{tab:size_and_kpoints}, 
where we report the value of $\delta U_{\rm th}$ per atom, i.e.
the quantity $\delta U_{\rm th} / N \equiv
[ \langle U_{\rm AI} - U_{\rm pair} \rangle_{\rm AI} -
U_{\rm th}^0 ] / N$ (see Sec.~\ref{sec:techniques}). Since $U_{\rm th}^0 / N$
is independent of the system size, the variation of the
reported quantity arises solely from size dependence of
$\langle U_{\rm AI} - U_{\rm pair} \rangle_{\rm AI} / N$.
We see that with $\sim 125$ atoms $\delta U_{\rm th} / N$
is converged to better than
5~meV/atom, and that already with 67 atoms the size error is of the
order of 10~meV/atom.

We tested for $k$-point errors in $\delta U_{\rm th}$ by 
performing calculations
using both four and 32 Monkhorst-Pack~\cite{monkhorst76} sampling
points. Since explicit AIMD calculations with so many $k$-points would
be extremely expensive, we use the following procedure. From an
existing $\Gamma$-point simulation we take a set of typically 10
atomic configurations separated by 0.1~ps. The {\em ab initio} total
energies of these configurations calculated with the different
$k$-point samplings are then compared. For sampling with 
four $k$-points,
we did calculations on systems of up to 241 atoms, but the heavier
calculations with 32 $k$-points were done only on the 67-atom system.
The results of these tests for the thermodynamic state
$V = 8.67$~\AA$^3$/atom and $T = 4300$~K
are also reported in
table~\ref{tab:size_and_kpoints}, where we see that for the
smallest system containing 67 atoms the difference with respect to a
calculation with the $\Gamma$-point only is $\sim 9$~meV/atom, but as
the number of atoms in the cell is increased above $\sim 125$ the
difference becomes negligible.  The result for the calculation with 32
$k$-points is identical to the one with four $k$-points and is not
reported in the table. We also found that 
fluctuations of the energy differences
between the calculations done with the $\Gamma$-point only and those with
4 $k$-points are extremely small.

Similar, but less extensive, tests of system-size and $k$-point
errors have also been performed at the state $V = 6.97$~\AA$^3$/atom,
$T = 6000$~K, and we find that the variation of these errors
with system size is numerically almost the same as before. The indication 
is therefore that $\delta U_{\rm th}$ can be obtained to a precision of 
{\em ca.}~5~meV/atom from simulations on systems of 125 atoms or more.
Unfortunately, it is not practicable yet to do all our AIMD
simulations with this system size, and in practice we have computed
$\delta U_{\rm th}$ from $\Gamma$-point simulations on the
67-atom system, and corrected the results by adding 10~meV/atom,
which from the present evidence appears to the almost constant error in
the $\Gamma$-point 67-atom results.

As expected, the numerical values of $\delta U_{\rm th}$
are small, and depend weakly 
on temperature and pressure across the range
of thermodynamic states of interest. We find that they
can be represented to within $\sim 3$~meV/atom by a sum of
third-degree polynomials in $V$ and $T$:
\begin{equation}
\delta U_{\rm th} / N = \sum_{i=0}^{3} \left( a_i V^i + 
b_i T^i \right) \; ,
\end{equation}
with the following fitting parameters (units of eV, \AA\ and K):
$a_0 = 0.649$;
$a_1 = -4.33 \times 10^{-2}$;
$a_2 = -4.19 \times 10^{-3}$;
$a_3 =  6.48 \times 10^{-5}$; 
$b_0 =  0.296$;
$b_1 = -6.51 \times 10^{-5}$;
$b_2 =  7.46 \times 10^{-9}$;
$b_3 = -2.07 \times 10^{-13}$.

To test the validity of neglecting the higher-order fluctuation
terms omitted from Eqn~(\ref{eqn:F_AI}),
we have performed full thermodynamic integration for
four different thermodynamic states, the first three with 
$V = 8.67$~\AA$^3$/atom and
$T = 4300$, 6000 and 8000~K and the fourth with 
$V = 6.97$~\AA$^3$/atom and $T = 8000$~K, using
the five equally spaced $\lambda$ values 0, 0.25, 0.5, 0.75 and
1.0. These calculations were done using $\Gamma$-point sampling on the
system of 67 atoms.  We have seen that this system size is not big
enough to yield the required precision for $F_{\rm AI}$, but it should
certainly be enough to test the adequacy of the second-order formula.
In Table~\ref{tab:small_lambda} we report a comparison between the results
obtained from the integral using the five $\lambda$ values and those from
the second order formula, and we see that they are practically
indistinguishable.
The Table also indicates that
the term $\langle ( \delta \Delta U )^2 \rangle_{\rm AI} \left/
2 k_{\rm B} T \right.$ 
is rather insensitive to thermodynamic state and can be
approximated to the required precision by setting it equal
to 10~meV/atom. We have used this constant value in 
evaluating the {\em ab initio} free energy by Eqn~(\ref{eqn:F_AI}). 

\section{Melting properties}
\label{sec:melting}
\begin{figure}
\psfig{figure=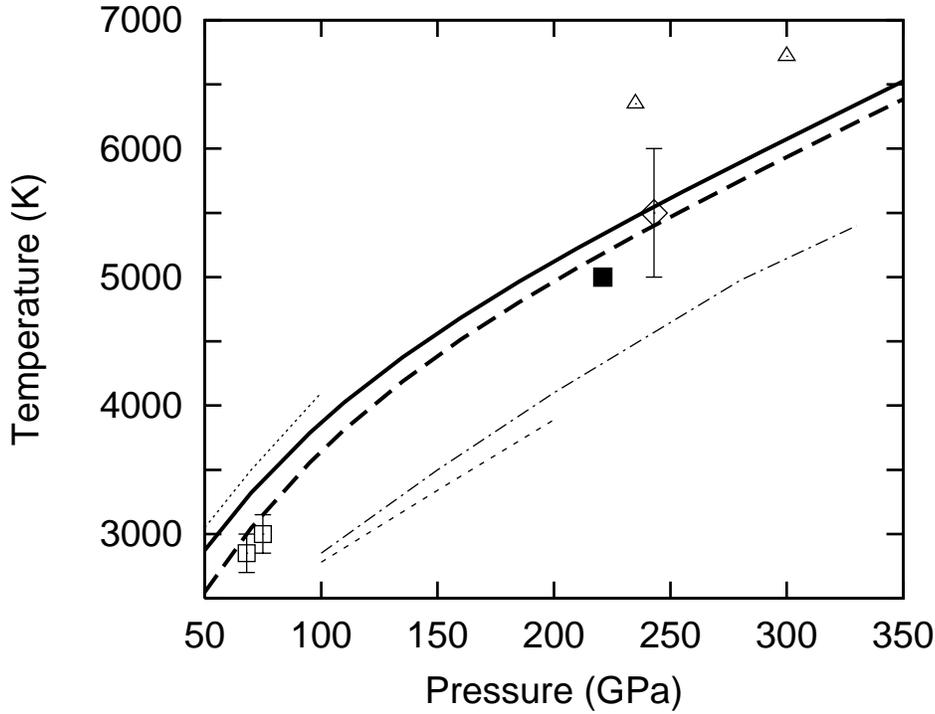}
\caption{Comparison of melting curve of Fe from present calculations
with previous experimental and {\em ab initio} results: heavy solid
and dashed curves: present work without and with free-energy
correction (see text); chain curve: {\em ab initio} results of
Ref.~\protect\cite{laio00}; dots, light dashes and squares: DAC
measurements of
Refs.~\protect\cite{williams87},~\protect\cite{boehler93}
and~\protect\cite{shen98}; triangles, diamond and solid square: shock
experiments of Refs.~\protect\cite{yoo93},~\protect\cite{brown86}
and~\protect\cite{holmes01}. Error bars are those quoted in original
references.}
\label{fig:melting_curve}
\end{figure}

From our parameterized formulas for the {\em ab initio} Helmholtz free
energies $F ( V , T )$ of the h.c.p. solid (Ref.~\cite{alfe01}) and
the liquid (present work), we immediately obtain the Gibbs free
energies $G ( p , T ) \equiv F ( V , T ) + p V$, and for each pressure
the melting temperature $T_m$ is determined as the $T$ at which the
latter free energies are equal for the solid and liquid. The resulting
melting curve is reported in 
Fig.~\ref{fig:melting_curve} for pressures from 50 to
350~GPa. On the same plot, we show the {\em ab initio} melting curve
reported very recently by Laio {\em et al.}~\cite{laio00}. We also
compare with experimental melting curves or points obtained by shock
experiments or by static-compression using the diamond anvil cell
(DAC).  DAC determinations of the melting curve of Fe and other
transition metals have been performed by several research
groups~\cite{boehler93,saxena94,shen98,errandonea01}.  The early results of
Williams {\em et al.}~\cite{williams87} lie considerably above those
of other groups, and are now generally discounted. This still leaves a
range of {\em ca.}~400~K in the experimental $T_{\rm m}$ at
100~GPa. Even allowing for this uncertainty, we acknowledge that our
melting curve lies appreciably above the surviving DAC curves, with
our $T_{\rm m}$ being above that of Shen {\em et al.}~\cite{shen98} by
{\em ca.}~400~K at 100~GPa. We return to this discrepancy below.

Shock measurements should in principle be able to fix a point on the
high-pressure melting curve at the thermodynamic state where melting
first occurs on the Hugoniot. However, temperature is notoriously
difficult to measure in shock experiments. The temperatures obtained
by Yoo {\em et al.}~\cite{yoo93} using pyrometric techniques are
generally regarded as being too high by at least 1000~K. This has been
confirmed by our recent {\em ab initio} calculations~\cite{alfe01} of
Hugoniot temperature for h.c.p. Fe. We therefore disregard their data
point on the melting curve. In the shock measurements of Brown and
McQueen~\cite{brown86} and Nguyen and Holmes~\cite{holmes01}, no
attempt was made to measure temperature, which was estimated using
models for the specific heat and Gr\"{u}neisen parameter; the
approximate validity of these models is supported by our {\em ab
initio} calculations~\cite{alfe01} on h.c.p.  Fe.  However, the
identification of the Hugoniot melting point has been hampered by the
possible existence of a solid-solid transition. In their measurements
of sound velocity on the Hugoniot, Brown and McQueen~\cite{brown86}
believed that they had observed a solid-solid transition as well as a
separate melting transition. The new shock results of Nguyen and
Holmes~\cite{holmes01} using improved techniques indicate that there
is no solid-solid transition, and we place greater weight on their
Hugoniot melting point.  We plot in Fig.~\ref{fig:melting_curve} the
point reported by Brown and McQueen~\cite{brown86} as lying on the
melting curve, though for the reasons just explained, we are cautious
about accepting it. We also plot the point obtained from the
measurements of Nguyen and Holmes~\cite{holmes01}. The pressure of 221
GPa is taken directly from their measurement of the onset of melting,
while the temperature at this point is taken from our calculation of
the Hugoniot temperature of the h.c.p. solid at this pressure, as
reported in Ref.~\cite{alfe01} (see also following section).


We now consider possible sources of error in our DFT
calculations. First, we recall that even with the best available GGA
for exchange-correlation energy the low-temperature $p(V)$ relation
for h.c.p. Fe is not in perfect agreement with experiment. This has been
shown by a number of independent calculations using all-electron
techniques~\cite{stixrude94,soderlind96} as well as
pseudopotential~\cite{vocadlo97} and PAW~\cite{alfe99b,kresse99}
techniques, all of which agree closely with each other. Roughly
speaking, the pressure is underpredicted by {\em ca.}~10~GPa at
near-ambient pressures and by {\em ca.}~8~GPa in the region of
300~GPa.  The pressure error can be thought of as arising from an error
in the Helmholtz free energy, so that the true free energy $F_{\rm
true}$ can be written as $F_{\rm true} = F_{\rm GGA} + \delta F$,
where $F_{\rm GGA}$ is our calculated free energy and $\delta F$ is
the correction. If we take the pressure error $\delta p \equiv - (
\partial \delta F / \partial V )_T$ to be linear in the volume, then
$\delta F$ can be represented as $\delta F = b_1 V + b_2 V^2$, where
$b_1$ and $b_2$ are adjustable parameters determined by least-squares
fitting to the experimental pressure. If we now neglect the
temperature dependence of $\delta F$, and simply add $\delta F ( V )$
to the calculated free energies of solid and liquid, this gives a way
of gauging our likely errors.  We find that this free-energy
correction leads to a lowering of the melting curve by {\em ca.}~350~K
in the region of 50~GPa and by {\em ca.}~70~K in the region of
300~GPa.

The second error source we consider is the PAW implementation,
and specifically our choice of the division
into core and valence states, and the PAW core radii.
As mentioned earlier, at
Earth's core pressures the $3p$ electrons, and to a lesser extent the
$3s$ electrons, must be treated as valence states. Moreover, the
choice of the maximum PAW core radius may also affect the
calculations, because under such high pressures and temperatures the
atoms come so close that the cores overlap. These errors may affect
the melting curve if they fail to cancel between the liquid and the
solid.  To check both these possible problems, we have performed
trial PAW calculations
with the much smaller core radius of 0.85~\AA\
and with both $3s$ and $3p$ states in the valence set; 
with this choice of core radius the
overlap of the cores in the liquid and the high temperature solid is
almost negligible. We have then used Eq.~(\ref{eqn:second_order}) to
calculate the free energy difference between the systems described
with the two PAW approximations, repeating the calculations for
both the liquid and the solid. To do that we have drawn two sets of 30
statistically independent configurations from two long simulations
performed with the original PAW approximation on the solid 
and the liquid at $V = 7.18$~\AA$^3$/atom and 
$T = 6700$~K. As expected, we find a significant shift in
the total electronic (free) energies. This shift is almost constant,
thus validating the use of Eq.~(\ref{eqn:second_order}), but the
important result is that it is almost the same for the liquid and the
solid, the two numbers being 
$F^l_{\rm hard} - F^l_{\rm soft} = -0.210$~eV/atom
$F^s_{\rm hard} - F^s_{\rm soft} = -0.204$~eV/atom. Here,
$F^l_{\rm hard}$ is the free energy calculated with small
core and $3s$ and $3p$ states in valence, and $F^l_{\rm soft}$ the free
energy with large core and the $3s$ and $3p$ frozen
in the core, plus the effective pair-potential; the superscripts $s$
and $l$ indicate the solid and the liquid respectively. The effect is
small, and stabilises the liquid by 6~meV/atom, which has the effect
of shifting the melting curve down by $\sim 60$~K.

As we show in Fig.~\ref{fig:melting_curve}, 
if we include both these corrections they bring
our low-temperature melting curve into quite respectable agreement
with the DAC measurements of Shen {\em et al.}, while leaving the
agreement with the shock points of Nguyen and Holmes essentially
unaffected. There is still a considerable discrepancy with the DAC
curve of Boehler~\cite{boehler93} and the {\em ab initio} results of
Laio {\em et al.}~\cite{laio00}

\begin{figure}
\psfig{figure=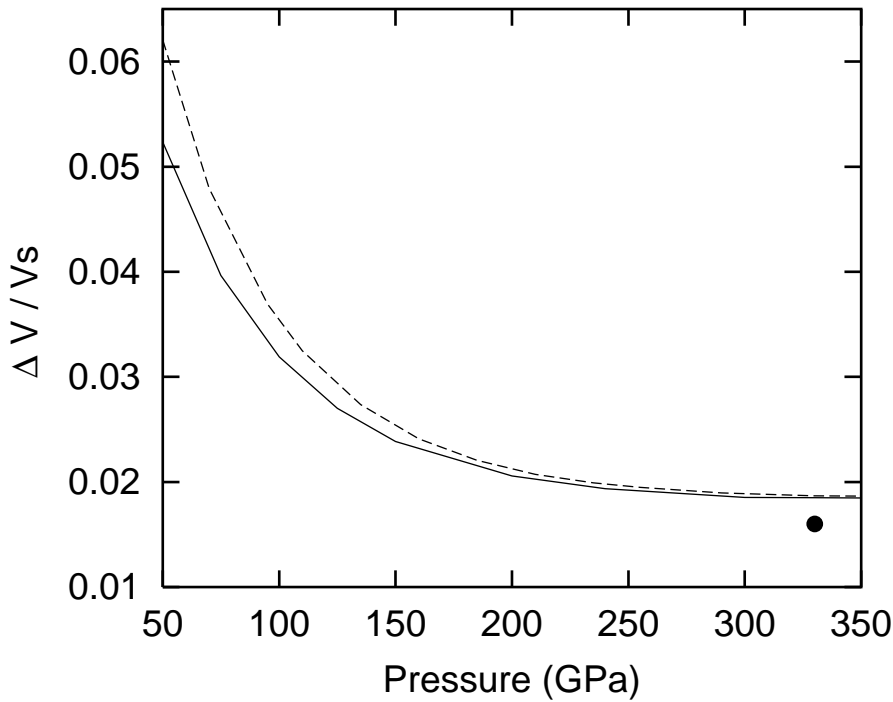}
\caption{{\em Ab initio} fractional volume change on melting of Fe as
a function of pressure. Solid and dashed curves: present work,
without and with free-energy correction (see text); black dot:
Ref.~\protect\cite{laio00}.}
\label{fig:melting_volume}
\end{figure}

We now turn to the changes of volume and entropy on melting.
Our calculated volume of melting (volume of liquid minus volume
of coexisting h.c.p. solid at each pressure expressed as a percentage
of the volume of the solid at that point) is plotted as a function of
pressure in Fig.~\ref{fig:melting_volume}. 
We also show the melting volume predicted by
the {\em ab initio} calculations of Laio {\em et al.}~\cite{laio00}
at the pressure 330~GPa, and it is encouraging to note that
their value of 1.6~\% is quite close to ours. The free-energy
correction discussed above makes only a small difference
to the calculated volume of melting: at 50~GPa the
correction makes the volume of melting increase from 5.0 to 5.8~\%,
while at 300~GPa it is affected by less than 0.1~\%.
The most striking feature of our results is the steep
decrease of $\Delta V$ by a factor of about three in the range
from 50 to 200~GPa, and its approximate constancy after that.

\begin{figure}
\psfig{figure=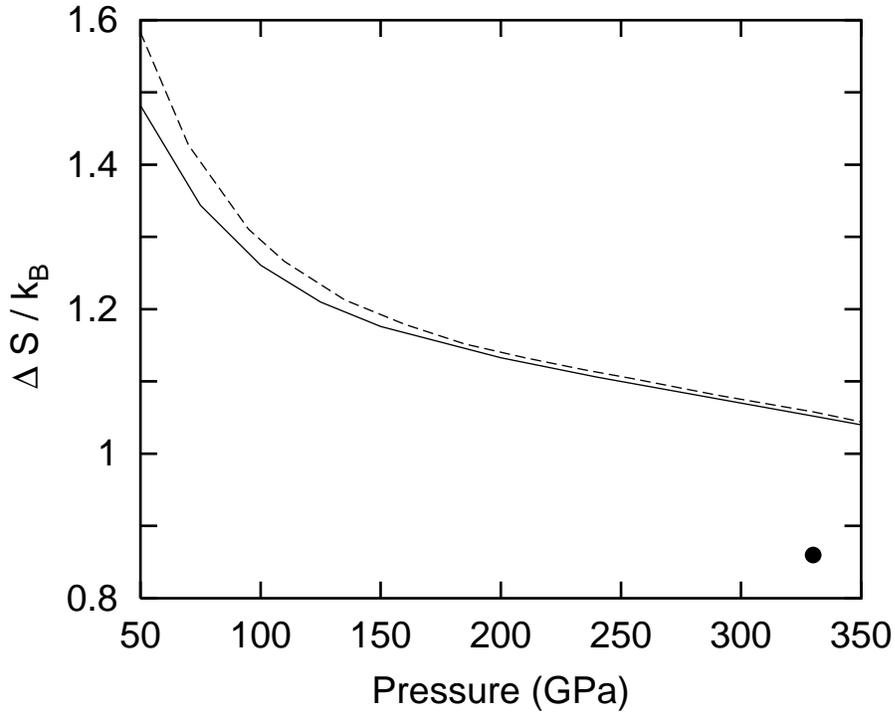}
\caption{{\em Ab initio} entropy change on melting per atom
(units of Boltzmann's constant $k_{\rm B}$). Solid and
dashed curves: present work, without and with free-energy
correction (see text); black dot: Ref.~\protect\cite{laio00}.}
\label{fig:melting_entropy}
\end{figure}

Our predicted entropy of melting $\Delta S_m$ (entropy per atom of
liquid minus entropy per atom of coexisting solid) is plotted as a
function of pressure in Fig.~\ref{fig:melting_entropy}, 
where we also show the {\em ab
initio} value of Laio~{\em et al.}~\cite{laio00} at 330~GPa. The
agreement of our value (1.05~$k_{\rm B}$) with theirs (0.86~$k_{\rm
B}$) is reasonably close. The entropy of melting also
decreases with increasing $p$, but more moderately than 
$\Delta V / V$, the decrease between 50 and 200~GPa being
only 30~\%. We note the relevance to the
slope of the melting curve, given by the Clausius-Clapeyron
relation: $d T_m / d p = \Delta V / \Delta S$. (This relation
is satisfied identically by our results, since they are
all derived from free energies.) The strong decrease of
$d T_m / d p$ between 50 and 200~GPa and its approximate
constancy thereafter is mainly due to the variation
of $\Delta V / V$.

\section{Hugoniot properties}
\label{sec:hugoniot}

Since shock experiments are the only direct way of obtaining thermodynamic
information for high-$p$/high-$T$ liquid Fe, it is important
to test our predictions against the available shock data. The data that
emerge most directly from shock experiments consist of a relation
between the pressure $p_{\rm H}$ and the molar volume $V_{\rm H}$
on the so-called Hugoniot line, which is the set of
thermodynamic states given by the Rankine-Hugoniot formula~\cite{poirier91}:
\begin{equation}
\frac{1}{2} p_{\rm H} ( V_0 - V_{\rm H} ) = E_{\rm H} - E_0 \; ,
\end{equation}
where $E_{\rm H}$ is the molar internal energy behind the shock front,
and $E_0$ and $V_0$ are the molar internal energy and volume in the
zero-pressure state ahead of the front.  The pressure-volume and
temperature-pressure relations on the Hugoniot are straightforwardly
obtained from our {\em ab initio} calculations: for a given $V_{\rm
H}$, one seeks the temperature $T_{\rm H}$ at which the
Rankine-Hugoniot relation is satisfied, and from this one obtains
$p_{\rm H}$ (and, if required, $E_{\rm H}$).  In experiments on Fe,
$V_0$ and $E_0$ refer to the zero-pressure b.c.c.  crystal. We obtain
$E_0$ directly from GGA calculations that we performed on the
ferromagnetic b.c.c. crystal, as described earlier~\cite{alfe01}, but
we use the experimental value of $V_0$. The slight shift produced by
using instead the theoretical value of $V_0$ was noted
earlier~\cite{alfe01}. Melting in shock experiments is usually
detected by monitoring the sound 
velocity~\cite{yoo93,brown86}, which
shows marked discontinuities of slope along the Hugoniot.  In a simple
melting transition, there are discontinuities at two characteristic
pressures $p_s$ and $p_l$, which are the points where the solid and
liquid Hugoniots meet the melting curve. Below $p_s$, the material
behind the shock front is entirely solid, while above $p_l$ it is
entirely liquid; between $p_s$ and $p_l$, the material is a two-phase
mixture.

\begin{figure}
\psfig{figure=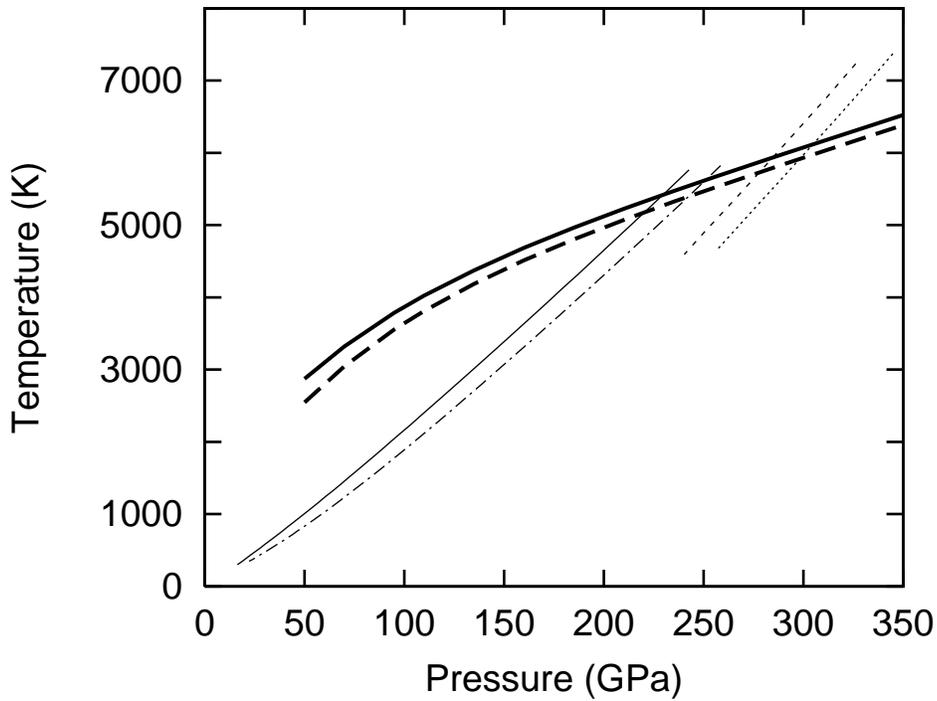}
\caption{Relation between {\em ab initio} melting curve
and {\em ab initio} Hugoniot temperature-pressure curves.
Heavy continuous and dashed curves: melting curves calculated
without and with free-energy correction (see text); light continuous
and chain curves: Hugoniot of solid without and with free-energy
correction; light dashed and dotted curves: Hugoniot of liquid
without and with free-energy correction.}
\label{fig:hugoniot_melting}
\end{figure}

We present in Fig.~\ref{fig:hugoniot_melting} 
our calculated $T_{\rm H} ( p_{\rm H} )$
Hugoniot curve for the liquid, together with our curve for the solid
reported earlier~\cite{alfe01} and our {\em ab initio} melting curve. Without
the free energy correction $\delta F$ of Sec.~\ref{sec:melting}, we find
$p_s =$~229 and $p_l =$~285~GPa. The very recent shock data
of Nguyen and Holmes~\cite{holmes01} give values 
of 221 and 260~GPa respectively,
so that our $p_s$ value is very close to theirs, and our $p_l$
value is also not very different.

If the correction $\delta F$ is included in calculating the
melting curve, then for consistency it must be included also
in the solid and liquid Hugoniots. It is straightforward
to obtain the corrected $p_{\rm H}$ and $E_{\rm H}$
as a function of $V_{\rm H}$ for the two phases. But in the
Rankine-Hugoniot equation we also need $E_0$ for the b.c.c.
crystal, and this will be subject to a correction similar to
$\delta F$, but of
unknown size. To supply the missing information, we add
to the {\em ab initio} energy of b.c.c. Fe a correction
term $\delta F_{\rm bcc}$, which we represent as $c_1 + c_2 V$.
The constants $c_1$ and $c_2$ are fixed by requiring that the 
equilibrium volume of the b.c.c. crystal and the low-temperature
transition pressure between the b.c.c. and h.c.p. phases be correctly
given. The resulting `corrected' $T_{\rm H} ( p_{\rm H} )$ Hugoniots
of the solid and liquid are reported in 
Fig.~\ref{fig:hugoniot_melting}. The shifts
in the curves are of about the same size as 
those discussed earlier~\cite{alfe01}
for the solid when we replaced the calculated b.c.c. volume
$V_0$ in the Rankine-Hugoniot equation by its experimental
value, and are an indication of the inherent uncertainty due
to DFT errors. The corrected values $p_s =$~243, $p_l =$~298~GPa
are now in somewhat poorer agreement with the experimental values. This
is a rather sensitive test of DFT errors, since the shallow angle
at which the Hugoniot curves cross the melting line amplifies
the effect of the errors.

\begin{figure}
\psfig{figure=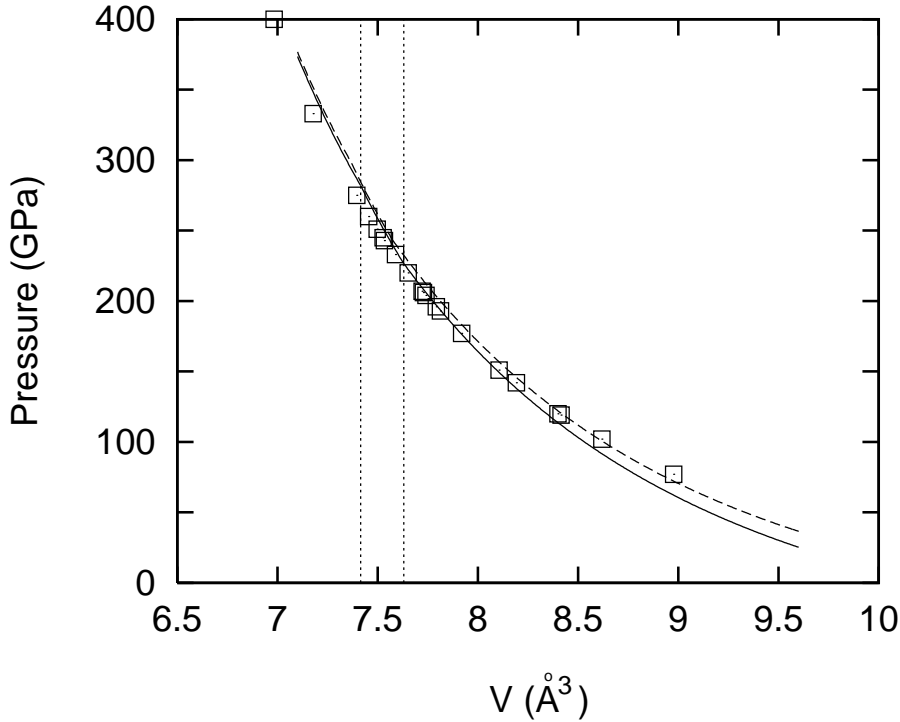,height=3.9in}
\caption{{\em Ab initio} Hugoniot pressure-volume curve
compared with experimental results of Ref.~\protect\cite{brown86}.
Solid and dashed curves: {\em ab initio} results without and with
free-energy correction (see text); squares: experimental
results. Vertical dotted lines indicate volumes at which melting
starts and finishes according to present (uncorrected) {\em ab initio}
results. To the right of rightmost vertical dotted line, curves represent
solid Hugoniot from Ref.~\protect\cite{alfe01}; to the left of leftmost
vertical line, curves represent present liquid Hugoniot.}
\label{fig:hugoniot}
\end{figure}

We now turn to our liquid-state results for $p_{\rm H} ( V_{\rm H} )$
compared with the shock data of Brown and McQueen~\cite{brown86}
(Fig.~\ref{fig:hugoniot}), including for completeness 
our results for the solid
reported earlier~\cite{alfe01}. We report results both with
and without the free energy correction $\delta F$, using the
experimental b.c.c. volume $V_0$ in the Rankine-Hugoniot
equation in both cases. We mark on the Figure the volumes
above which the shocked material is entirely solid
and below which it is entirely liquid.  Above the upper volume,
we report our calculated h.c.p. Hugoniots, and below the lower
volume the liquid Hugoniots. In the interval between them,
we linearly mix the two. We note that the $\delta F$
correction makes little difference to the liquid
Hugoniot, which lies above the experimental values
by {\em ca.}~3~\%.

\begin{figure}
\psfig{figure=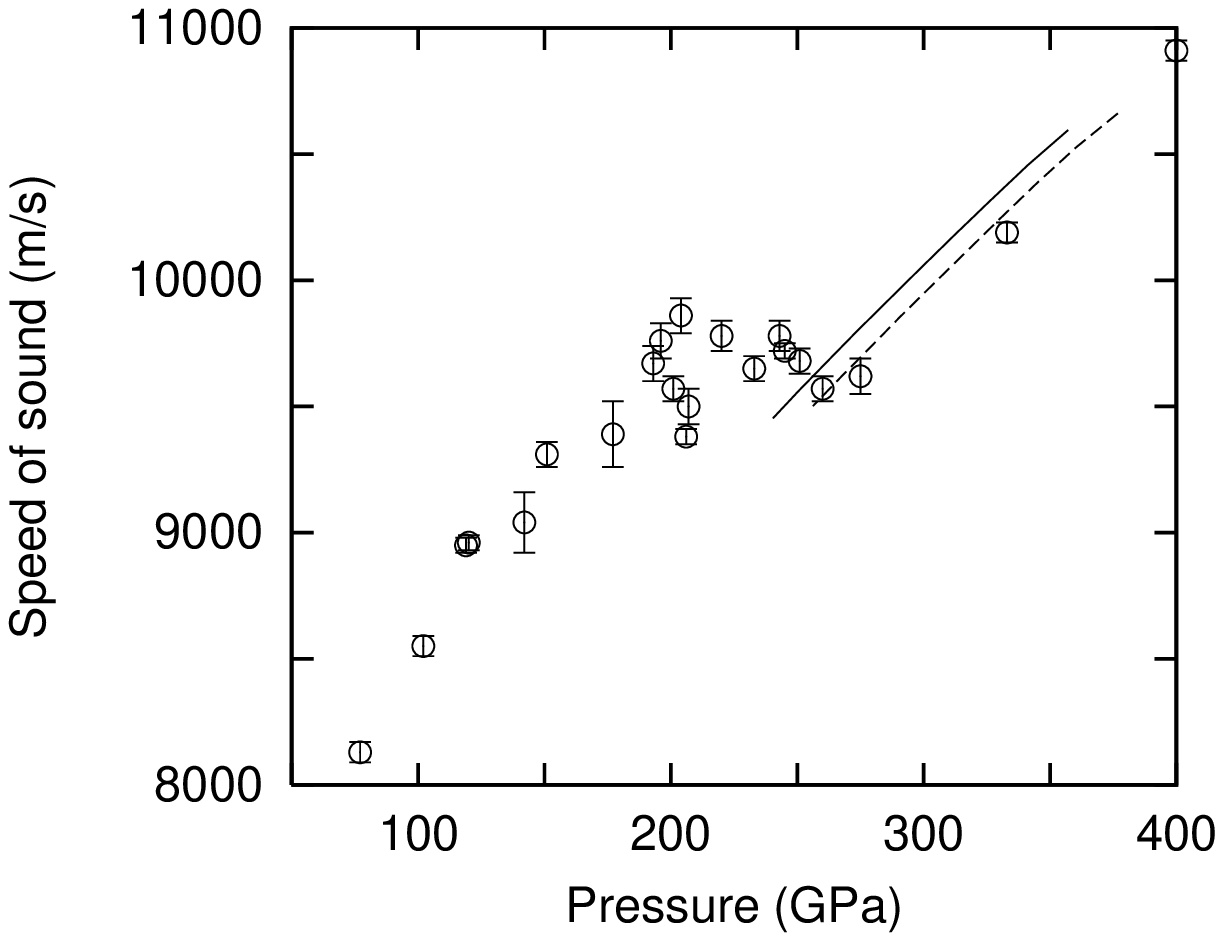,height=3.9in}
\caption{Longitudinal speed of sound on the Hugoniot. Circles:
experimental values from Ref.~\protect\cite{brown86}; continuous and
dashed curves: present {\em ab initio} values without and with
free-energy correction (see text).}
\label{fig:speed_of_sound}
\end{figure}

Shock experiments on Fe have given values for the adiabatic sound
speed $v_S = ( K_S / \rho )^{1/2}$ of the liquid, with $K_S$ the
adiabatic bulk modulus and $\rho$ the mass density. 
Fig.~\ref{fig:speed_of_sound} shows our
{\em ab initio} values for $v_S$ of the liquid as a 
function of pressure on the
Hugoniot, both with and without the $\delta F$ correction, compared
with the shock data of Refs.~\cite{brown86}.
Up to the pressure of $\sim 260$~GPa, the experimental
points refer to the solid or
the two-phase region, so it is in the liquid region above this
pressure that the comparison is significant. In that region,
our agreement with the experimental data is close, the
discrepancies being $\sim 2$ and $< 1$~\% for our uncorrected
and corrected $v_S$ values respectively.

We conclude this Section by reporting results for the Gr\"{u}neisen
parameter $\gamma$ on the liquid Hugoniot. This parameter
is defined as $\gamma \equiv V ( \partial p / \partial E )_V =
\alpha K_T V_{\rm m} / C_v$, with $\alpha$ the volume expansion
coefficient, $K_T$ the isothermal bulk modulus, $C_v$ the
constant-volume molar specific heat, and $V_{\rm m}$ the
molar volume. Assumptions or estimates of its values
have played a key role in constructing parameterised equations
of state for Fe. Our calculated $\gamma$ on the liquid Hugoniot
is almost exactly constant, varying in the narrow range from
1.51 to 1.52 as $p$ goes from 280 to 340~GPa.

\section{Thermodynamics of the liquid}
\label{sec:liq_therm}

Although directly measured data on high-$p$/high-$T$ liquid Fe all
come from shock experiments, attempts have been made to combine
these data with measurements at lower $p$ and $T$ using
parameterised models for quantities such as $K_S$, $\gamma$ and
$C_v$ to estimate thermodynamic properties away from the
Hugoniot curve~\cite{anderson94}. These attempts have been crucial in trying to
understand how the properties of the Earth's liquid core deviate
from those of pure liquid Fe. We present here a brief comparison
with these experimentally based extrapolations for the two quantities
that determine the seismic properties of the outer core: the density
$\rho$ and the adiabatic bulk modulus $K_S$. 

Since the outer core is in a state of turbulent convection, the
variation of its thermodynamic properties with depth is expected to
follow an adiabat. We therefore present our comparisons on adiabats
specified by their temperature $T_{\rm ICB}$ at $p = 330$~GPa, which
is the pressure at the inner-core/outer-core boundary
(ICB)~\cite{prem}. We choose the two temperatures $T_{\rm ICB} = 5000$
and 7000~K, because the results of Sec.~\ref{sec:melting} indicate
that the melting temperature at the ICB pressure lies between these
limits. Our comparisons (Tables~\ref{tab:rho} and \ref{tab:K_S}) 
show that the uncorrected {\em ab initio} density is very close
(within a few tenths of a percent) to the extrapolated
experimental data at $p = 150$~GPa, and in slightly poorer
agreement (within $\sim 1.5$~\%) at $p = 350$~GPa. As expected,
the free-energy correction lowers the predicted density,
resulting in larger discrepancies with experiment of 1.5~\%
and 2.5~\% at $p = 150$ and 350~GPa respectively. Our uncorrected
{\em ab initio} $K_S$ values also agree more closely with the
experimental data, being typically within 2~\%, while the
corrected predictions disagree with the data by up to 8~\%.
However, given the closer agreement between {\em ab initio}
and experiment on the Hugoniot (Sec.~\ref{sec:hugoniot}), it
is possible that some of the disagreements may be due to deficiencies
in the experimental extrapolation.

\section{Discussion and conclusions}
\label{sec:discussion}

In assessing the reliability of our results, we consider three sources
of error: first, the uncontrolled DFT errors inherent in the GGA for
exchange and correlation energy; second, the controllable errors in
the detailed electronic-structure implementation of GGA, and
specifically in the use of PAW to calculate the total {\em ab initio}
(free) energy $U_{\rm AI} ( {\bf R}_1 , \ldots {\bf R}_N ; T_{\rm el}
)$ for each set of atomic positions ${\bf R}_1 , \ldots {\bf R}_N$;
third, the statistical-mechanical errors, including system-size
effects. We have endeavoured to reduce errors of the third kind below
10~meV/atom for the liquid. In our earlier free-energy calculations on
the h.c.p. solid~\cite{alfe01}, the corresponding error was estimated as $\sim
15$~meV/atom. Taking these errors together, and recalling that the
resulting error in melting temperature $T_m$ is roughly the combined
free-energy error divided by Boltzmann's constant, 
we find an expected $T_m$
error of {\em ca.}~$\pm 300$~K.  We have also attempted to control
errors of the second kind by changing the division between
core and valence states and by
reducing the core radius. These tests suggest that the associated
error in $T_m$ is probably no more than {\em ca.}~$\pm 100$~K. The
inherent DFT errors are more difficult to quantify, but we have
demonstrated that the known discrepancies in the low-temperature
$p(V)$ relation for h.c.p. Fe almost certainly lead to an overestimate
of $T_m$ by {\em ca.}~350~K at 50~GPa and {\em ca.}~70~K at 300~GPa,
and we have corrected for this. We have also seen the significant
shifts in the Hugoniot curves resulting from DFT errors. We believe
the remaining uncertainty in $T_m$ from this source could be as much
as 300~K.

Our attempts to correct for DFT errors give a melting curve which is
in quite good agreement with the recent measurements of Shen {\em et
al.}~\cite{shen98}, and with estimates based on shock
data~\cite{brown86}; the methods used to estimate temperature in
the shock experiments are also supported by our {\em ab initio}
results for Gr\"{u}neisen parameter $\gamma$ and specific heat
$C_v$~\cite{alfe01}. Our melting curve is still above the experimental data of
Boehler~\cite{boehler93} by $\sim 800$~K in the pressure region up to
{\em ca.~}~100~GPa.  We cannot rule out the possibility that some
of this
discrepancy is due to our DFT errors. Our substantial disagreement
with the {\em ab initio} melting curve of Laio {\em et
al.}~\cite{laio00} must be due to other reasons.
We are currently working with authors of Ref.~\cite{laio00} to
discover the cause of the disagreement, and we hope to report on this
in the future.

A key part of our strategy for eliminating system-size errors in the
calculated free energies is the use of an empirical reference model
which accurately reproduces the fluctuations of total energy. At first
sight, the use of a reference model based on a purely repulsive pair
potential might seem surprising, since it does not explicitly include
a description of metallic bonding.  An empirical reference model
(there called an `optimised potential model') is also used in the work
of Laio {\em et al.}~\cite{laio00}, though they use it in a different
way from us.  Their optimised potential model is a form of the
`embedded atom model' (EAM)~\cite{daw93,baskes92,belonoshko97}, which
explicitly includes metallic bonding. As described in our earlier
work~\cite{alfe01}, we have investigated the consequences 
of using the EAM as a
reference model. We showed there that for present purposes
fluctuations of the bonding energy are negligible, and that 
under these circumstances the EAM is
almost exactly equivalent to a model based on repulsive pair
potentials. We also showed that there is no numerical advantage in
using the EAM as reference model for the calculation of free
energies. The use of different reference models {\em per se} therefore
appears to have nothing to do with the current disagreement between
{\em ab initio} melting curves.

The agreement of our {\em ab initio} results with the limited data
from shock experiments on the liquid is reasonably satisfactory.  In
particular, our predicted Hugoniot relation $p_{\rm H} ( V_{\rm H} )$
is almost as good as we found earlier for the solid. The adiabatic
sound velocity of the liquid is also predicted to within $1 - 3$~\%,
the discrepancy depending on whether or not we attempt to correct for
DFT errors. The good agreement for the Gr\"{u}neisen parameter
$\gamma$ is also encouraging. Our results for the
h.c.p. solid~\cite{alfe01} indicated that $\gamma$ varies little with
pressure or temperature for $100 < p < 300$~GPa and $4000 < T <
6000$~K, and has a value of {\em ca.}~1.5. Our present results
indicate that the same is true of the liquid.

The {\em ab initio} free-energy techniques outlined here could
clearly be adapted to a wide range of other problems, so that
melting curves could be calculated for many materials, including
those of geological interest, like silicates. We have recently
completed {\em ab initio} calculations of the melting curve
of aluminium up to pressures of 150~GPa, which are
in excellent agreement with static-compression and shock data, as
will be reported elsewhere~\cite{vocadlo01}.

In conclusion, we have shown how {\em ab initio} free-energy
calculations based on thermodynamic integration can be used
to obtain the melting curve and the volume and entropy of
melting of a material over a wide pressure range. We have
emphasised that the key requirement on the reference
system used in thermodynamic integration is that it faithfully mimics
the fluctuations of {\em ab initio} energy in thermal equilibrium.
Our {\em ab initio} melting curve of Fe over the pressure range
$50 - 350$~GPa agrees fairly well with experimental data obtained
from both static-compression and shock techniques, but
significant discrepancies remain to be resolved. Our {\em ab initio}
predictions for quantities obtained directly from shock
experiments, including the Gr\"{u}neisen parameter of the liquid,
agree closely with the measured data in most cases.

\section*{Acknowledgments}

The work of DA was supported by NERC Grant GST/02/1454 to
G.~D.~Price and M.~J.~Gillan, and by a 
Royal Society University Research Fellowship. We thank NERC and EPSRC
for allocations of time on the Cray T3E machines at Edinburgh
Parallel Computer Centre and Manchester CSAR service, these allocations
being provided by the Minerals Physics Consortium (GST/02/1002)
and U.K. Car-Parrinello Consortium (GR/M01753). Calculations
were also performed at the UCL HiPerSPACE Centre funded by
the Joint Research Equipment Initiative. We thank Dr.~G.~Kresse
for technical assistance with the PAW calculations.

\pagebreak

\pagebreak

\begin{table}
\begin{tabular}{l|ccccc}
 &  \multicolumn{5}{c}{$\rho$ (kg m$^{-3}$) } \\
\hline
 $T$ (K) & 9540 & 10700 & 11010  & 12130 & 13300 \\
\hline
  3000 & 0.097  &         &       &       &     \\  
       & (60)   &         &       &       &     \\
  4300 &        &  0.085  &       &       &     \\  
       &        &  (132)  &       &       &     \\
  5000 &        &  0.089  &       &       &     \\
       &        &  (140)  &       &       &     \\  
  6000 & 0.104  &  0.096  & 0.089 & 0.103 & 0.125 \\  
       & (90)   &  (151)  & (170) & (251) & (360) \\
  7000 &        &  0.093  & 0.098 & 0.109 & 0.131 \\ 
       &        &  (161)  & (181) & (264) & (375) \\ 
  8000 &        &  0.092  & 0.099 & 0.104 & 0.124 \\ 
       &        &  (172)  & (191) & (275) & (390) \\  
\end{tabular}
\caption{Normalised fluctuation strength $\sigma$ (see text)
characterising the accuracy with which the inverse-power
reference model mimics the energy fluctuations of
{\em ab initio} liquid Fe. Values of $\sigma$ (eV units) are
given for a set of AIMD simulations at different
densities $\rho$ and temperatures $T$. Pressure at each thermodynamic
state (GPa units) is given in parenthesis.}
\label{tab:fluct}
\end{table}

\hspace{0.5in}

\begin{table}
\begin{tabular}{l|ccc}
 & $( \delta U_{\rm th} ( N ) -
\delta U_{\rm th} ( 241 ) ) / N$ (eV)  & 
$( \delta U_{\rm th} ( 4 k ) - 
\delta U_{\rm th} ( \Gamma) ) / N$ (eV) &
$\sigma$ (eV) \\
\hline
67   & -0.009 $\pm$ 0.002 & 0.009 $\pm 0.003$ & 0.085 \\
89   & -0.012 $\pm$ 0.001 & 0.007 $\pm 0.002$ & 0.073 \\
107  & -0.010 $\pm$ 0.001 & 0.006 $\pm 0.002$ & 0.083 \\
127  &  0.004 $\pm$ 0.001 & 0.000 $\pm 0.002$ & 0.086 \\
157  &  0.001 $\pm$ 0.001 & 0.001 $\pm 0.002$ & 0.069 \\
199  &  0.001 $\pm$ 0.001 &                   & 0.082 \\
241  &  0.000 $\pm$ 0.001 & 0.001 $\pm 0.002$ & 0.101 \\
\end{tabular}
\caption{Dependence on number of atoms $N$ in the simulation
cell of size errors and $k$-point sampling errors in the
quantity $\delta U_{\rm th}$ entering the {\em ab initio}
free energy of liquid Fe (see Eqn~(\ref{eqn:F_AI})). Second column reports
$\delta U_{\rm th} / N$ (eV units) with a constant offset
chosen so that the reported value for the largest system size
is zero. Third column reports difference of $\delta U_{\rm th} / N$
between simulations using four $k$-points and $\Gamma$-point
sampling. Fourth column reports normalised fluctuation
strength $\sigma$ (see text) for different system sizes.}
\label{tab:size_and_kpoints}
\end{table}

\hspace{0.5in}

\begin{table}
\begin{tabular}{l|cccc}
 $T$ (K) & $N^{-1} \int_0^1 d \lambda \,
\langle \Delta U \rangle_\lambda$ (eV) &
$\langle ( \delta \Delta U )^2 \rangle_{\rm AI} / 2 N k_{\rm B} T$ (eV) \\
\hline
  4300 & 0.012  &  0.012 \\  
  6000 & 0.010  &  0.009 \\ 
  8000 & 0.006 (0.010) &  0.006 (0.010) \\ 
\end{tabular}
\caption{Difference $F_{\rm AI} - F_{\rm ref}$ between
free energies of {\em ab initio} and reference systems
calculated in two ways: by full thermodynamic integration
(column 2) as in Eqn~(\ref{eqn:thermint_3}), and by the second-order
fluctuation approximation (column 3) as in Eqn~(\ref{eqn:2nd_order}). Free
energy differences are given per atom in eV units for
three temperatures at the density $\rho = 10700$~kg~m$^{-3}$.
Values in parenthesis refer to the density $\rho = 13300$~kg~m$^{-3}$.}
\label{tab:small_lambda}
\end{table}

\begin{table}
\begin{tabular}{l|cccccc}
 & \multicolumn{4}{c}{$\rho$ (kg m$^{-3}$) }\\
 & \multicolumn{2}{c}{This work} & \multicolumn{2}{c}{Experiments} \\ 
\hline
 $P$~(GPa) &  $T = 5000$~K & $T = 7000$~K  &
$T = 5000$~K & $T = 7000$~K \\
\hline 
150 & 11075 (10930) & 10806 (10659) & 11110 & 10800 \\
200 & 11738 (11625) & 11477 (11350) & 11870 & 11560 \\
250 & 12323 (12220) & 12059 (11950) & 12440 & 12180 \\
300 & 12844 (12756) & 12575 (12481) & 13000 & 12800 \\
350 & 13315 (13232) & 13043 (12970) & 13550 & 13290 \\
\end{tabular}
\caption{Comparison of {\em ab initio} and experimental
density $\rho$ of liquid Fe on two adiabats, with
adiabats specified by the temperature $T$ at the pressure
$p = 300$~GPa. {\em Ab initio} $\rho$ values are given
both without and with (in parenthesis) free-energy correction
$\delta F$ (see text).}
\label{tab:rho}
\end{table}

\begin{table}
\begin{tabular}{l|cccccc}
 & \multicolumn{4}{c}{$K_S$ (GPa) }\\
 & \multicolumn{2}{c}{This work} & \multicolumn{2}{c}{Experiments} \\ 
\hline
$P$ (GPa) &  $T = 5000$~K & $T = 7000$~K  &
$T = 5000$~K & $T = 7000$~K \\
\hline 
150 & 708 (662) & 656 (613) & 695 & 668 \\
200 & 878 (838) & 820 (781) & 877 & 849 \\
250 & 1050 (1010) & 981 (944) & 1058 & 1016 \\
300 & 1220 (1180) & 1140 (1103) & 1232 & 1193 \\
350 & 1384 (1350) & 1296 (1264) & 1400 & 1355 \\
\end{tabular}
\caption{Comparison of {\em ab initio} and experimental
adiabatic bulk modulus $K_S$ of liquid Fe on two adiabats,
with adiabats specified by the temperature $T$ at
the pressure $p = 330$~GPa. {\em Ab initio} $K_S$ values
are given both without and with (in parenthesis) free-energy
correction $\delta F$ (see text).}
\label{tab:K_S}
\end{table}


\begin{thebibliography}{99}

\bibitem{sugino95}
O. Sugino and R. Car, Phys. Rev. Lett. {\bf 74}, 1823 (1995).

\bibitem{smargiassi95a}
E. Smargiassi, P. A. Madden, Phys Rev B {\bf 51}, 117 (1995).

\bibitem{dewijs98}
G. A. de Wijs, G. Kresse and M. J. Gillan, Phys. Rev. B {\bf 57},
8223 (1998).

\bibitem{alfe99a}
D. Alf\`{e}, G. A. de Wijs, G. Kresse and M. J. Gillan,
Int. J. Quant. Chem., {\bf 77}, 871 (2000). 

\bibitem{karki00} B. B. Karki, R. M. Wentzcovitch, S. de Gironcoli,
S. Baroni, Phys. Rev. B, {\bf 62}, 14750 (2000).

\bibitem{Lichtenstein00} A. I. Lichtenstein, R. O. Jones, S. de
Gironcoli, S. Baroni, Phys. Rev. B {\bf 62},  11487 (2000).

\bibitem{xie99a}
J. J. Xie, S. P. Chen, J. S. Tse, S. de Gironcoli, S.  Baroni,
Phys. Rev. B, {\bf 60}, 9444 (1999).

\bibitem{xie99b}
J. J. Xie, S. de Gironcoli, S.  Baroni, M. Scheffler,
Phys. Rev. B, {\bf 59}, 965 (1999).

\bibitem{lazzeri98}
M. Lazzeri, S. de Gironcoli, Phys. Rev. Lett., {\bf 81}, 2096 (1998).

\bibitem{pavone98}
P. Pavone, S. de Gironcoli, S. Baroni,  Phys. Rev. B, {\bf 57}, 10421 (1998).

\bibitem{alfe01}
D. Alf\`e, G. D. Price, M. J. Gillan, Phys. Rev. B, {\bf 64}, 045123 (2001).

\bibitem{alfe99d}
D. Alf\`e,  M. J. Gillan and G. D. Price, Nature, {\bf 401}, 462 (1999).

\bibitem{birch64} F. Birch, J. Geophys. Res., {\bf 69}, 4377 (1964).

\bibitem{ringwood77} A. E. Ringwood, Geochem. J., {\bf 11}, 111
(1977).

\bibitem{poirier94} J.-P. Poirier, Phys. Earth Planet. Inter. {\bf
85}, 319 (1994).

\bibitem{anderson97}
For a brief review of experiments on the melting curve of Fe, see
e.g. O. L. Anderson and A. Duba, J. Geophys. Res. Sol. Earth,
{\bf 102}, 22659 (1997).

\bibitem{boehler93}
R. Boehler, Nature {\bf 363}, 534 (1993).

\bibitem{saxena94}
S. K. Saxena, G. Shen, and P. Lazor, Science, {\bf 264}, 405 (1994). 

\bibitem{shen98}
G. Shen, H. Mao, R. J. Hemley, T. S. Duffy and M. L. Rivers,
Geophys. Res. Lett. {\bf 25}, 373 (1998).

\bibitem{errandonea01}
D. Errandonea, B. Schwager, R. Ditz, C. Gessmann, R. Boehler,
M. Ross, Phys. Rev. B {\bf 63}, 132104 (2001).

\bibitem{williams87}
Q. Williams, R. Jeanloz, J. D. Bass, B. Svendesen, T. J. Ahrens, Science {\bf 286}, 181
(1987).

\bibitem{yoo93}
C. S. Yoo, N. C. Holmes, M. Ross, D. J. Webb and C. Pike,
Phys. Rev. Lett. {\bf 70}, 3931 (1993).

\bibitem{brown86}
J. M. Brown and R. G. McQueen, J. Geophys. Res. {\bf 91},
7485 (1986).

\bibitem{holmes01}
J. H. Nguyen and N. C. Holmes, unpublished.

\bibitem{laio00}
A. Laio, S. Bernard, G. L. Chiarotti, S. Scandolo and E. Tosatti,
Science {\bf 287}, 1027 (2000).

\bibitem{belonoshko00} A. B. Belonoshko, R. Ahuja, and B. Johansson,
Phys. Rev. Lett. {\bf 84}, 3638 (2000).

\bibitem{frenkel96}
D. Frenkel and B. Smit, {\em Understanding Molecular Simulation},
Academic Press, San Diego (1996).

\bibitem{wang91}
Y. Wang and J. Perdew, Phys. Rev. B {\bf 44}, 13298 (1991).

\bibitem{perdew92}
J. P. Perdew, J. A. Chevary, S. H. Vosko, K. A. Jackson,
M. R. Pederson, D. J. Singh and C. Fiolhais, Phys. Rev. B
{\bf 46}, 6671 (1992).

\bibitem{stixrude94}
L. Stixrude, R. E. Cohen and D. J. Singh, Phys. Rev. B {\bf 50}, 6442 (1994).

\bibitem{soderlind96}
P. S\"{o}derlind, J. A. Moriarty and J. M. Willis, Phys. Rev. B
{\bf 53}, 14063 (1996).

\bibitem{vocadlo97}
L. Vo\v{c}adlo, G. A. de Wijs, G. Kresse, M. J. Gillan and G. D. Price,
Faraday Disc. {\bf 106}, 205 (1997).

\bibitem{alfe99b}
D. Alf\`{e}, G. Kresse and M. J. Gillan, Phys. Rev. B, {\bf 61}, 132 (2000).

\bibitem{blochl94}
P. E. Bl\"{o}chl, Phys. Rev. B {\bf 50}, 17953 (1994).

\bibitem{kresse99}
G. Kresse and D. Joubert, Phys. Rev. B {\bf 59}, 1758 (1999).

\bibitem{wei85}
S. H. Wei and H. Krakauer, Phys. Rev. Lett., {\bf 55},
1200 (1985).

\bibitem{vanderbilt90}
D. Vanderbilt, Phys. Rev. B {\bf 41}, 7892 (1990).

\bibitem{kresse96a}
G. Kresse and J. Furthm\"{u}ller, Phys. Rev. B {\bf 54},
11169 (1996).

\bibitem{kresse96b}
G. Kresse and J. Furthm\"{u}ller, Comput. Mater. Sci. {\bf 6},
15 (1996).

\bibitem{alfe99e}
D. Alf\`{e}, Comp. Phys. Commun., {\bf 118}, 31 (1999).

\bibitem{mermin65}
N. D. Mermin, Phys. Rev. {\bf 137}, A1441 (1965).

\bibitem{laird92}
B. B. Laird and A. D. J. Haymet, Mol. Phys., {\bf 75}, 71 (1992).

\bibitem{johnson93} K. Johnson, J. A. Zollweg, and E. Gubbins,
Mol. Phy. {\bf 78}, 591 (1993).

\bibitem{monkhorst76} 
H. J. Monkhorst and J. D. Pack, Phys. Rev. B {\bf 13}, 5188 (1976).

\bibitem{poirier91}
J.-P. Poirier, {\em Introduction to the Physics of the Earth's
Interior}, Cambridge University Press, Cambridge (1991), ch.~4.

\bibitem{anderson94}
W. W. Anderson and A. T. J. Ahrens, 
J. Geophys. Res. {\bf 99}, 4273 (1994).

\bibitem{prem} A. M. Dziewonski and D. L. Anderson, Phys. Earth
Planet. Inter. {\bf 25}, 297 (1981).

\bibitem{daw93}
M. S. Daw, S. M. Foiles, and M. I. Baskes, Mat. Sci. Rep., {\bf 9}, 251 (1993).

\bibitem{baskes92}
M. I. Baskes, Phys. Rev. B, {\bf 46}, 2727 (1992).

\bibitem{belonoshko97}
A. B. Belonoshko, and R. Ahuja,
Phys. Earth Planet. Inter., {\bf 102}, 171 (1997).

\bibitem{vocadlo01}
L. Vo\v{c}adlo and D. Alf\`e, unpublished.


\end{thebibliography}
\end{document}